\documentclass[
floatfix,
aps,
preprint,
superscriptaddress,
showpacs,
preprintnumbers,
amsmath,assume,
prb,
longbibliography,
]{revtex4-1}
\usepackage{graphicx}
\usepackage{hyperref}
\usepackage[mathlines]{lineno}
\usepackage[utf8]{inputenc}
\usepackage{epstopdf, epsfig}
\usepackage{mathrsfs}
\usepackage{xcolor}
\usepackage{amsmath}
\usepackage{abraces}
\usepackage[utf8]{inputenc}

\begin{document}

\preprint{APS/123-QED}

\title{Non-resonant effects in pilot-wave hydrodynamics}

\author{Bauyrzhan K. Primkulov}
\thanks{contributed equally}
\affiliation{Department of Mathematics, Massachusetts Institute of Technology, Cambridge, MA, USA}%
\affiliation{Department of Mechanical Engineering, Yale University, New Haven, CT, USA}%

\author{Davis J. Evans}
\thanks{contributed equally}
\affiliation{Department of Mathematics, Massachusetts Institute of Technology, Cambridge, MA, USA}%

\author{Joel B. Been}
\affiliation{Department of Mathematics, Massachusetts Institute of Technology, Cambridge, MA, USA}%

\author{John W.M. Bush}%
\email{bush@math.mit.edu}
\affiliation{Department of Mathematics, Massachusetts Institute of Technology, Cambridge, MA, USA}%

\date{\today}

\begin{abstract}
Pilot-wave hydrodynamics concerns the dynamics of 'walkers,' droplets walking on a vibrating bath, and has provided the basis for the burgeoning field of hydrodynamic quantum analogs. We here explore a theoretical model of pilot-wave hydrodynamics that relaxes the simplifying assumption of resonance between the droplet and its pilot wave, specifically the assumption of a fixed impact phase between the bouncing drop and its wave field. The model captures both the vertical and horizontal dynamics of the drop, allowing one to examine non-resonant effects for both free and constrained walkers. The model provides new rationale for a number of previously reported but poorly understood features of free walker motion in pilot-wave hydrodynamics, including colinear swaying at the onset of motion, intermittent walking, and chaotic speed oscillations, all of which are accompanied by sporadic changes in the impact phase of the bouncing drop. The model also highlights the degeneracy in the droplets' vertical dynamics, specifically, the possibility of two distinct bouncing phases and of switching between the two. Consideration of this degeneracy is critical to understanding the droplet dynamics and statistics emerging in confined geometries at high memory and the interaction of walking droplets with standing Faraday waves.
\end{abstract}

\maketitle


\section{Introduction}\label{sec:intro}
The hydrodynamic pilot-wave system, wherein a drop is propelled by the wavefield it generates by bouncing on a vibrating fluid bath, has broadened the scope of classical mechanics to encompass features once believed to be exclusive to quantum systems~\citep{bush_new_2015, bush_pilot-wave_2015, bush_introduction_2018, bush_hydrodynamic_2021}. The drop and the wavefield form a single entity known as a ‘walker,’ and the resulting physical picture is reminiscent of de Broglie’s proposed quantum-scale particle dynamics~\citep{de_broglie_recherches_1924}. In light of the longstanding conceptual difficulties in quantum mechanics~\citep{bricmont_history_2017, kay_escape_2024}, this similarity has inspired an extensive exploration of hydrodynamic quantum analogs (HQAs) over the past two decades~\citep{bush_new_2015, bush_pilot-wave_2015, bush_introduction_2018, bush_hydrodynamic_2021} and has concurrently sparked a revisitation of de Broglie's pilot-wave mechanics~\citep{drezet_mechanical_2020, dagan_hydrodynamic_2020, durey_hydrodynamic_2020, darrow_revisiting_2024}. Two HQAs are of particular interest to our study, the hydrodynamic analogs of a simple harmonic oscillator and the quantum corral. \citet{perrard_self-organization_2014} investigated the behaviour of walkers subject to the two-dimensional harmonic potential and found that a double-quantization of walker orbits emerges as one tunes the strength of the confining potential and the memory of the system. Specifically, periodic orbits arise that are quantized in both mean radius and mean angular momentum. In the hydrodynamic corral~\citep{harris_wavelike_2013, saenz_statistical_2017}, a droplet walks within a bounded domain for approximately an hour, after which the droplet's probability density function closely resembles that of electrons trapped in a quantum corral~\citep{crommie_confinement_1993, crommie_imaging_1993}. Our study of non-resonant effects will inform both of these canonical HQAs.

\begin{figure}
    \centering
    \includegraphics[width=\linewidth]{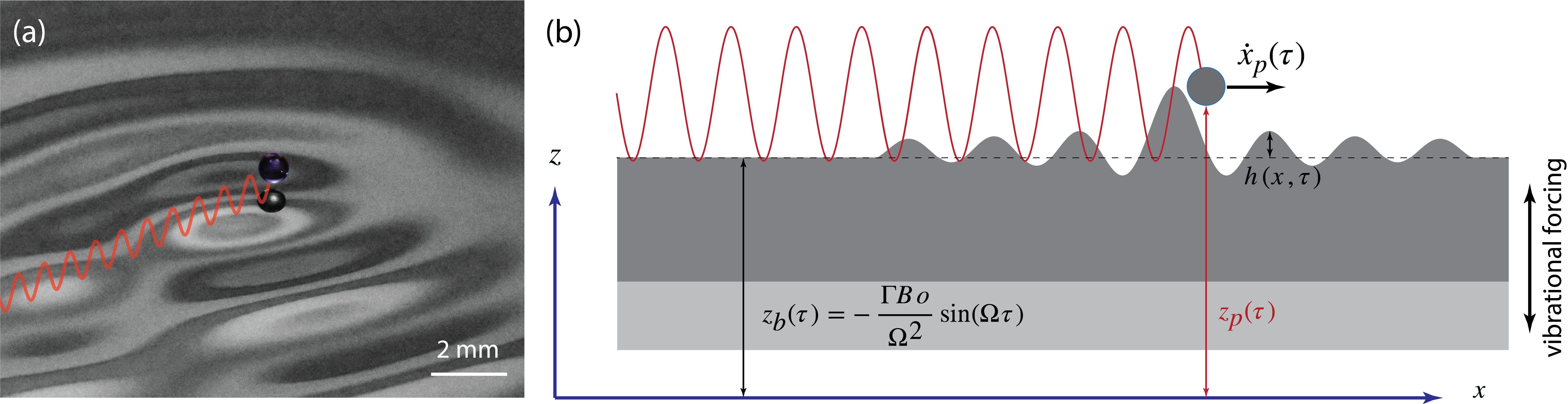}
    \caption{(a)~Photograph of a millimetric walker traversing a silicone oil bath. The walker is composed of both droplet and guiding (or 'pilot') wave. The red curve illustrates the drop's path. (b)~Schematic illustration of a walker bouncing on a liquid bath, showing the positions of the drop's base $z_p(\tau)$ and the unperturbed liquid surface $z_b(\tau)$ in the lab frame of reference. The bath is subject to vibrational forcing so that $z_b(\tau)=-\frac{\Gamma \mathit{Bo}}{\Omega^2}\sin(\Omega\tau)$. The pilot-wave height $h(x,\tau)$ is measured with respect to the unperturbed liquid surface.}
    \label{fig:schem}
\end{figure}

A liquid bath of silicone oil subjected to periodic vibrational acceleration $\gamma \sin\omega t$ destabilizes to a subharmonic field of Faraday waves of wavelength $\lambda_F$ and frequency $\omega_F=\omega/2$ when the Faraday threshold is exceeded, $\gamma>\gamma_F$~\citep{benjamin_stability_1954}. Below this threshold, $\gamma<\gamma_F$, a millimetric silicone oil drop can bounce indefinitely on the bath owing to the sustenance of an intervening air layer during impact~\citep{walker_amateur_1978, couder_bouncing_2005, terwagne_lifetime_2007, terwagne_metastable_2009}. When the drop bounces at or near the Faraday frequency, its impact on the bath surface generates a localized subharmonic quasi-monochromatic standing wave pattern with the Faraday wavelength $\lambda_F$ and frequency $\omega_F=\omega/2$. The longevity of these waves increases as the bath's vibrational acceleration approaches $\gamma_F$, allowing the bath to retain the ``memory'' of the droplet's history~\citep{eddi_information_2011}. One can colloquially think of the memory parameter $\mathit{Me}$ (defined in Table~\ref{tab:parameters}) as the number of prior bounces whose waves still influence the drop. \citet{couder_walking_2005} discovered that, in a narrow parameter regime, the wave may destabilize the bouncing droplet, transforming it into a walker that executes rectilinear horizontal motion at a constant speed $u_0$ (Fig.~\ref{fig:schem}). 

A hierarchy of theoretical models of increasing sophistication and complexity have been developed to capture different aspects of this pilot-wave hydrodynamic system~\citep{turton_review_2018, bush_hydrodynamic_2021}. Pilot-wave hydrodynamics has three characteristic timescales: (i)~the drop's bouncing timescale ($\tau_b=1/\omega_F\approx1/40$ sec), (ii)~the timescale of horizontal dynamics ($\tau_h=\lambda_F/u_0\approx 5$ sec), and (iii)~the timescale of statistical convergence ($\tau_\text{stat}\approx1$ hour). The level of detail accounted for in a given model is determined by the time scale of interest. One can loosely classify the existing models into full and reduced models. \citet{molacek_drops_2013a, molacek_drops_2013b} developed the first full model of the drop dynamics, both vertical and horizontal, but employed a reduced model for the wave dynamics. \citet{milewski_faraday_2015} developed the first full wave model using the weakly viscous potential flow equations and adopted Moláček and Bush's~\citep{molacek_drops_2013b} model for the drop dynamics, specifically the logarithmic spring model for the vertical dynamics. Subsequently, \citet{galeano-rios_non-wetting_2017, galeano-rios_ratcheting_2018} replaced the logarithmic spring force during the drop-bath impacts by kinematic matching conditions between the droplet, modelled as a non-wetting rigid sphere, and the liquid bath. While the latter two models are the most detailed to date, they require meshing of the fluid bath domain and resolving the partial differential equations for the wave height at every grid point, which is computationally intensive. Consequently, these full models are typically limited to simulating relatively short experiments ($\tau_b \lesssim \tau \lesssim \tau_h$)~\citep{galeano-rios_ratcheting_2018}, and so have not been used to resolve the emergent statistics ($\tau \ll \tau_\text{stat}$).

\begin{figure}
    \centering
    \includegraphics[width=0.9\linewidth]{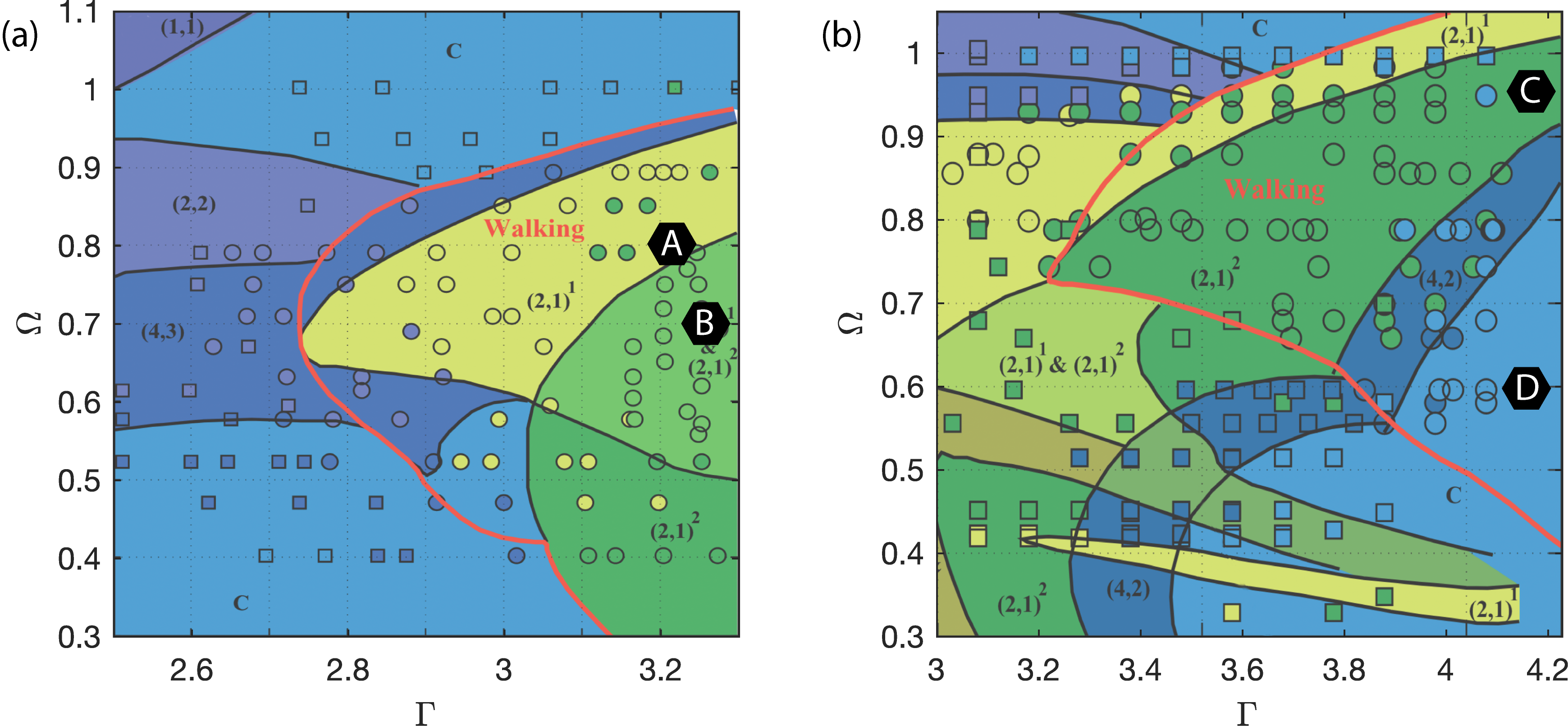}
    \caption{Phase diagram adapted from \citet{wind-willassen_exotic_2013}, that shows the bouncing and walking modes for $20$\;cSt silicone oil driven at (a)~$f=70$\;Hz and (b)~$f=80$\;Hz. Here, $\Gamma=\gamma/g$ is the dimensionless vibrational acceleration of the bath, and $\Omega=\omega/\omega_D$ is the vibration number---the ratio of the bath's angular frequency, $\omega=2\pi f$, and the drop's characteristic natural frequency $\omega_D=\sqrt{{\sigma}/{\rho R^3}}$. In the $(m,n)^i$ mode, the drop's bouncing motion is periodic over $m$ vibrational forcing periods, during which it undergoes $n$ impacts. The subscript $i$ ranks various $(m, n)$ states according to their total mechanical energy. For example, resonant walkers may assume either $(2,1)^1$ or $(2,1)^2$ modes, the latter being more energetic. Experimental data is marked with squares for bouncers and with circles for walkers. Black hexagons in the phase diagrams correspond to $(\Omega,\Gamma)$ values used in our simulations.}
    \label{fig:phase_diag}
\end{figure}

Stroboscopic models are based on the assumption of resonance between drop and wave~\citep{oza_trajectory_2013} as arises in the majority of the walking regime (Fig.~\ref{fig:phase_diag}). When such resonance is achieved, time-averaging over the period of the drop's vertical motion effectively eliminates the vertical dynamics from consideration. These models adopt the reduced wave description of \citet{molacek_drops_2013b} and are computationally efficient in that they only require computation of the wave field directly beneath the drop~\citep{durey_classical_2021}. The computational efficiency of the stroboscopic models allows them to span the timescales of horizontal dynamics and statistical convergence ($\tau_h \lesssim \tau \lesssim \tau_\text{stat}$). These stroboscopic models have been successful in rationalizing many of the experimental observations, including the emergence of quantized orbits in a rotating frame~\citep{oza_pilot-wave_2014-1, liu_pilot-wave_2023} and a central force~\citep{labousse_pilot-wave_2016}. Moreover, they have served as a foundation for the generalized pilot-wave framework~\citep{bush_pilot-wave_2015}, which allows for a numerical exploration of classical pilot-wave dynamics in a parameter space inaccessible in the laboratory~\citep{turton_review_2018, durey_speed_2020}. Stroboscopic models are known to have shortcomings in rationalizing certain aspects of pilot-wave hydrodynamics, including walker pair interactions~\citep{galeano-rios_ratcheting_2018, oza_orbiting_2017, arbelaiz_promenading_2018}, where non-resonant effects are known to be important. With a view to addressing these shortcomings, Couchman et al.~\citep{couchman_bouncing_2019, couchman_free_2020} developed a modified semi-empirical stroboscopic model that first approximated the effects of phase variations, and so illustrated the influence of these variations on the stability of droplet pairs~\citep{couchman_bouncing_2019} and rings~\citep{couchman_free_2020}. Most notably, the stroboscopic models have proven to be inadequate in rationalizing the emergent quantum-like statistics in corrals~\citep{durey_faraday_2020}, where non-resonant effects are known to be significant~\citep{harris_wavelike_2013, saenz_statistical_2017}.

We here relax the assumption of drop-wave resonance in order to capture non-resonant effects. The resulting model resolves both horizontal and vertical dynamics of the drop, and is sufficiently efficient to capture non-resonant walking modes across all three characteristic time scales of pilot-wave hydrodynamics ($\tau_b \lesssim \tau \lesssim \tau_\text{stat}$). In \S\ref{sec:experimental}, we highlight the experiments, both performed in this study and previously reported, that provide evidence of non-resonant features. In \S\ref{sec:model}, we present the details of our non-resonant model. In \S\ref{sec:free_walkers}, we use our model to rationalize a range of non-resonant effects observed with free walkers, including the emergence of degenerate bouncing phases, the swaying onset of rectilinear motion, as well as mode-switching, intermittent, and chaotic walkers. In \S\ref{sec:confined}, we discuss the non-resonant behaviour of droplets walking in confinement, specifically the sporadic flips in impact phase at high memory. Finally in \S\ref{sec:conclusions}, we discuss the applications of our model to systems that are above the Faraday threshold and marked by stochastic phase switching.

\begin{figure}
    \centering
    \includegraphics[width=\linewidth]{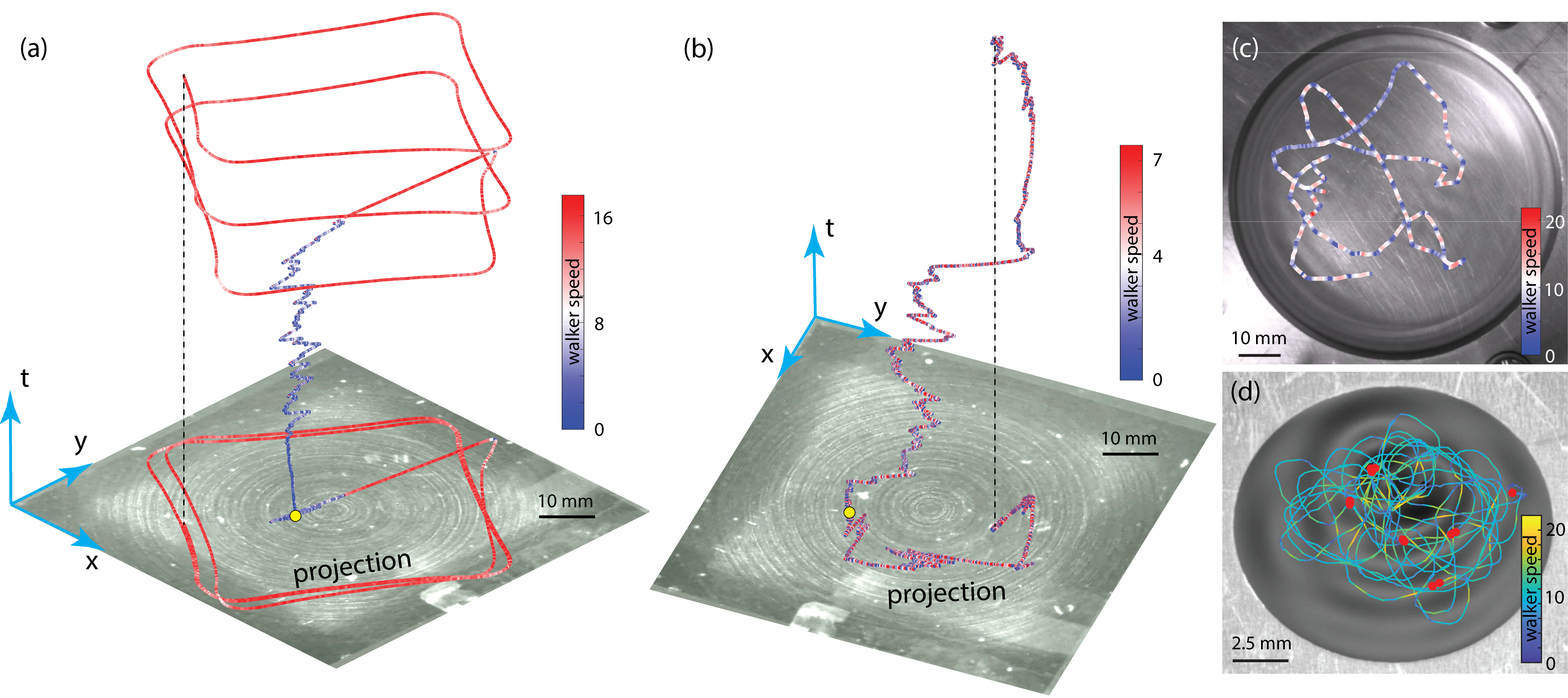}
    \caption{Non-resonant effects in pilot-wave hydrodynamics. (a)~The swaying onset of rectilinear motion, which is most pronounced for relatively large drops at $f=70$\;Hz (Video 1). The experiment corresponds to point A in Fig.~\ref{fig:phase_diag}a. (b)~An intermittent walker at $\Omega=0.9$, $f=80$\;Hz, and $\Gamma/\Gamma_F=0.95$ (Video 2), marked by intermittent changes in walking direction. The experiment corresponds to point C in Fig.~\ref{fig:phase_diag}b. Intermittent walkers tend to move sporadically along a straight line but occasionally switch direction, likely due to ambient air currents. In (a) and (b), experimental trajectories are plotted in $(x,y,t)$ space and color-coded by walker speed; the same trajectories are then projected onto the $(x,y)$ plane. (c)~A ``mixed-state'' walker, as reported by \citet{wind-willassen_exotic_2013} (Video 3), marked by periodic in-line oscillations. The experiment corresponds to a region near point B in Fig.~\ref{fig:phase_diag}a. (d)~Flipping of the walker's impact phase evident in the experiments of \citet{saenz_statistical_2017} (Video 4), where red dots mark the locations of the phase flips.}
    \label{fig:experiments}
\end{figure}

\section{Experiments}\label{sec:experimental}
The most comprehensive experimental study of the droplet bouncing modes to date was conducted by \citet{wind-willassen_exotic_2013}. They examined how the system's behavior depends on drop size, vibrational forcing frequency and acceleration, and identified a variety of exotic bouncing modes, which they summarized in a phase diagram. In Fig.~2, we reproduce their phase diagrams for vibrational forcing at 70 and 80 Hz, the two frequencies considered in our study. In Fig.~2, the colored regions represent different bouncing modes, while the symbols indicate experimental data: squares correspond to bouncers, and circles to walkers. The red line indicates the walking threshold predicted by the logarithmic spring model of \citet{molacek_drops_2013b}.

A number of curious effects can be observed experimentally with the walking-droplet system but have yet to be rationalized theoretically. For example, we here show that walkers starting from rest may sway back and forth along a straight line before reaching their steady walking speed~(Fig.~\ref{fig:experiments}a). Moreover, one can find corners of the parameter regime (e.g., near point C in Fig.~\ref{fig:phase_diag}b) where walkers never reach a steady walking speed, but instead exhibit intermittent motion along a straight line with sporadic reversals in direction (Fig.~\ref{fig:experiments}b). These intermittent walkers have been alluded to, but not carefully characterized, by \citet{molacek_drops_2013b} and \citet{wind-willassen_exotic_2013}. Walkers can also exhibit spontaneous and persistent in-line oscillations in their walking speed (Fig.~\ref{fig:experiments}c). When confined to a small domain at high memory, walkers can lose their resonance with the wave field~\citep{saenz_statistical_2017}, resulting in sporadic flipping between two distinct walker states and a reversal in the walking direction that acts to erase the drop's pilot-wave field~(Fig.~\ref{fig:experiments}d)~\citep{perrard_wave-based_2016}. We here rationalize all such effects with our non-resonant walker model.

\section{Coupled Walker Model}\label{sec:model}
We adapt the model of \citet{molacek_drops_2013a, molacek_drops_2013b}; specifically, we adopt the linear spring model to describe the vertical dynamics of the drop, departing from the logarithmic spring model utilized in \citet{wind-willassen_exotic_2013}. The logarithmic spring model was originally introduced to reconcile differences between experimental observations at high $\mathit{Me}$ and theoretical predictions based on the assumption of an unperturbed free surface at drop impacts~\citep{molacek_drops_2013a, molacek_drops_2013b}. As highlighted by \citet{couchman_bouncing_2019}, while this assumption holds at low memory, it is invalid at high memory where a substantial wave field persists beneath the drop. We here take into account the form of the underlying time-dependent wavefield when calculating the force during drop-bath impacts, and so more accurately capture the two-way droplet-wave coupling. Finally, we find it simplest to describe the dynamics in the lab frame of reference, in which the liquid bath is driven sinusoidally (Fig.~\ref{fig:schem}).

We non-dimensionalize the drop's equations of motion using the drop radius $R$ as a characteristic length and the drop's natural period of oscillations $\omega_D^{-1}$ as the characteristic time (see Table~\ref{tab:parameters}). The resulting dimensionless governing equations thus take the form
\begin{align}
    \Ddot{z}_p(\tau)= \overbrace{F_N(\tau)}^{\text{impact force}}-\overbrace{\mathit{Bo}}^{\text{gravity}}, \label{eqn:zdot} \\
    \mathbf{\ddot{x}}_p(\tau) +\underbrace{(\mathcal{D}_h F_N(\tau) + \frac{9}{2}\mathit{Oh}_a) \mathbf{\dot{x}}_p(\tau)}_{\text{impact drag + air drag}} = - \underbrace{F_N(\tau) \nabla{h(\mathbf{x}_p,\tau)}}_{\text{horizontal impulses}}, \label{eqn:xdot}
\end{align}
where $\mathbf{x}_p(\tau)$ and $z_p(\tau)$ are the horizontal and vertical coordinates of the drop's base. Dimensionless numbers are defined in Table~\ref{tab:parameters}. Equation~\eqref{eqn:zdot} represents the vertical force balance on the drop, where Bond number $\mathit{Bo}$ represents the gravitational pull on the drop and $F_N(\tau)$ is the dimensionless normal force that the drop experiences during impact. This normal force is comprised of the linear spring and damping components:
\begin{equation}
    F_N(\tau)=-\mathcal{H}(-z_p+z_b+h) [\mathcal{D}_v (\dot{z}_p-\dot{z}_b-\dot{h})+\mathcal{C}_v ({z}_p-{z}_b-{h})], \label{eqn:FN}
\end{equation}
where $z_b(\tau)=-\frac{\Gamma \mathit{Bo}}{\Omega^2}\sin(\Omega\tau)$ is the position of the unperturbed fluid surface, $h(\mathbf{x}_p,\tau)$ is the instantaneous wave height perturbation from $z_b(\tau)$ (see Fig.~\ref{fig:schem}), $\mathcal{D}_v$ and $\mathcal{C}_v$ are damping and spring constants, and $\mathcal{H}(-z_p+z_b+h)$ is the Heaviside function that ensures that $F_N(\tau)$ is non-zero only during drop-bath impacts. Equation~\eqref{eqn:xdot} indicates that the horizontal motion of the drop is driven by a series of impulses proportional to the local slope of the wave field, and resisted by drag. The wave field is a linear superposition of standing waves generated by prior drop impacts~\citep{molacek_drops_2013b, turton_review_2018, couchman_bouncing_2019}:
\begin{equation}
    h(\mathbf{x},\tau) =  \cos{(\Omega \tau/2)} \sum_{i=1}^{n} A_i e^{-\frac{\tau-\tau_i}{\tau_F \mathit{Me}}} (\tau-\tau_i)^{-1/2} J_0 (k_F r) [1+(\xi r K_1(\xi r)-1)e^{-r^{-2}}],
    \label{eqn:h}
\end{equation}
where $r=|\mathbf{x}-\mathbf{x}_i|$, and $e^{-\frac{\tau-\tau_i}{\tau_F \mathit{Me}}} (\tau-\tau_i)^{-1/2}$ prescribes the temporal decay of the wave~\citep{molacek_drops_2013b}, and $J_0 (k_F r) [1+(\xi r K_1(\xi r)-1)e^{-r^{-2}}]$ is the wave kernel that accounts for the spatial damping~\citep{couchman_bouncing_2019}. The amplitude of each standing wave depends on the timing of the impact according to
\begin{equation*}
    A_i=\frac{4}{3}\sqrt{2\pi \mathit{Oh}_e}\frac{k_F^3}{3k_F^2+\mathit{Bo}} \int_{\tau_c} F_N(s) \sin{(\Omega s/2)} ds,
\end{equation*}
where $\tau_c$ is the total contact time. Note that for delta function impacts $F_N(s)=\delta(s-\tau_i)$, the amplitude $A_i\rightarrow0$ whenever the impact time $\tau_i$ satisfies $\Omega \tau_i/2=2\pi i$. Thus, the sign and amplitude of the standing wave necessarily depend on the phase of impact. For impacts of finite extent, impact times and locations are defined, respectively, by
\begin{equation*}
    \tau_{i} = \frac{\int_{\tau_c} F_N(s)sds}{\int_{\tau_c} F_N(s)ds}, \hspace{6mm}
    \mathbf{x}_{i} = \frac{\int_{\tau_c} F_N(s) \mathbf{x}_p(s) ds}{\int_{\tau_c} F_N(s)ds},
\end{equation*}
where the integrals are evaluated over the droplet's contact time $\tau_c$. 

The drop trajectory is now prescribed by the solution to the coupled ordinary differential equations~\eqref{eqn:zdot}-\eqref{eqn:xdot}, provided the wave field is updated after each impact. The narrow bounds for parameters $\mathcal{D}_h$, $\mathcal{D}_v$, $\mathcal{C}_v$ in the walking regime have been justified in \citet{molacek_drops_2013a,molacek_drops_2013b}, and the values used in our study are listed in Table~\ref{tab:parameters}, along with the remainder of the physical parameters. We use a fourth-order Runge Kutta method to integrate the vertical and horizontal dynamics, and treat the wave field as a forcing function. Each time the droplet is launched from the bath, the wavefield is updated by computing $\tau_i$ and $\mathbf{x}_i$ from the numerical data and updating $h(\mathbf{x},\tau)$ with a new Bessel function term in Eq.~\eqref{eqn:h}.

\begin{table}
\caption{Definition of relevant variables and parameters}
\label{tab:parameters}
\centering
 \begin{tabular}{p{4cm} p{12cm}} 
 \hline
 \emph{Symbol} & \emph{Definition} \\
 $\rho$, $\sigma$, $R$ & drop and bath density and surface tension, drop radius \\
 $\mu$, $\mu_a$ & dynamic viscosities of oil and air \\
 $f$, $\omega=2\pi f$ & driving frequency, angular driving frequency \\
 $\omega_D=\sqrt{\frac{\sigma}{\rho R^3}}$, $\Omega~=~\frac{\omega}{\omega_D}$ & drop's natural frequency, and vibration number \\
 $x_p$, $y_p$, $z_p$, $h$ &  drop coordinates, and wave height normalized by $R$ \\ 
 $k_F$ & dimensionless Faraday wavenumber\\ 
 $k_F^3+k_F\mathit{Bo} = \Omega^2/4$ & dispersion relation relevant for the deep-bath limit~\citep{benjamin_stability_1954} \\ 
 $\mu_e$, $\mathit{Oh}_e=\frac{\mu_e}{\sqrt{\sigma \rho R}}$ & effective viscosity and Ohnesorge number \citep{molacek_drops_2013b}\\
 $\mathit{Oh}_a=\frac{\mu_a}{\sqrt{\sigma \rho R}}$ & Ohnesorge number based on air viscosity\\
 $\mathit{Bo}=\frac{\rho g R^2}{\sigma}$ & Bond number \\
 $\tau=\omega_D t$, $\tau_F=\frac{4\pi}{\Omega}$ & time normalized by $\omega_D$, dimensionless Faraday period \\
 $\tau_d \approx \frac{1}{\mathit{Oh}_e k_c^2}$ & dimensionless wave decay time~\citep{molacek_drops_2013b}, where $\frac{k_c}{k_F} \approx 1-\epsilon^2$ and $\epsilon=\frac{\mathit{Oh}_e \Omega k_F}{3 k_F^2+Bo}$ \\
 $\Gamma=\frac{\gamma}{g}$, $\Gamma_F=\frac{\gamma_F}{g}$ & dimensionless driving acceleration and Faraday threshold \\
 $\mathit{Me} = \frac{\tau_d}{\tau_F(1-\Gamma/\Gamma_F)}$ & memory parameter \\
 $\xi = \sqrt{\frac{2\epsilon^2}{k_F^2 \mathit{Oh}_e \tau_F \mathit{Me} (1+2\epsilon^2)}}$ & dimensionless spatial damping coefficient~\citep{couchman_bouncing_2019} \\
 $\mathcal{D}_h=0.17$ & horizontal damping coefficients from impact~\citep{molacek_drops_2013b} \\
 $\mathcal{D}_v = 0.48$, $\mathcal{C}_v = 0.59$ & damping coefficient and spring constant in the vertical drop dynamics~\citep{couchman_bouncing_2019} \\
 \hline
 \end{tabular}
\end{table}

\section{Free walkers }\label{sec:free_walkers}
\subsection{Periodic free walkers}
\begin{figure}
    \centering
    \includegraphics[width=0.9\linewidth]{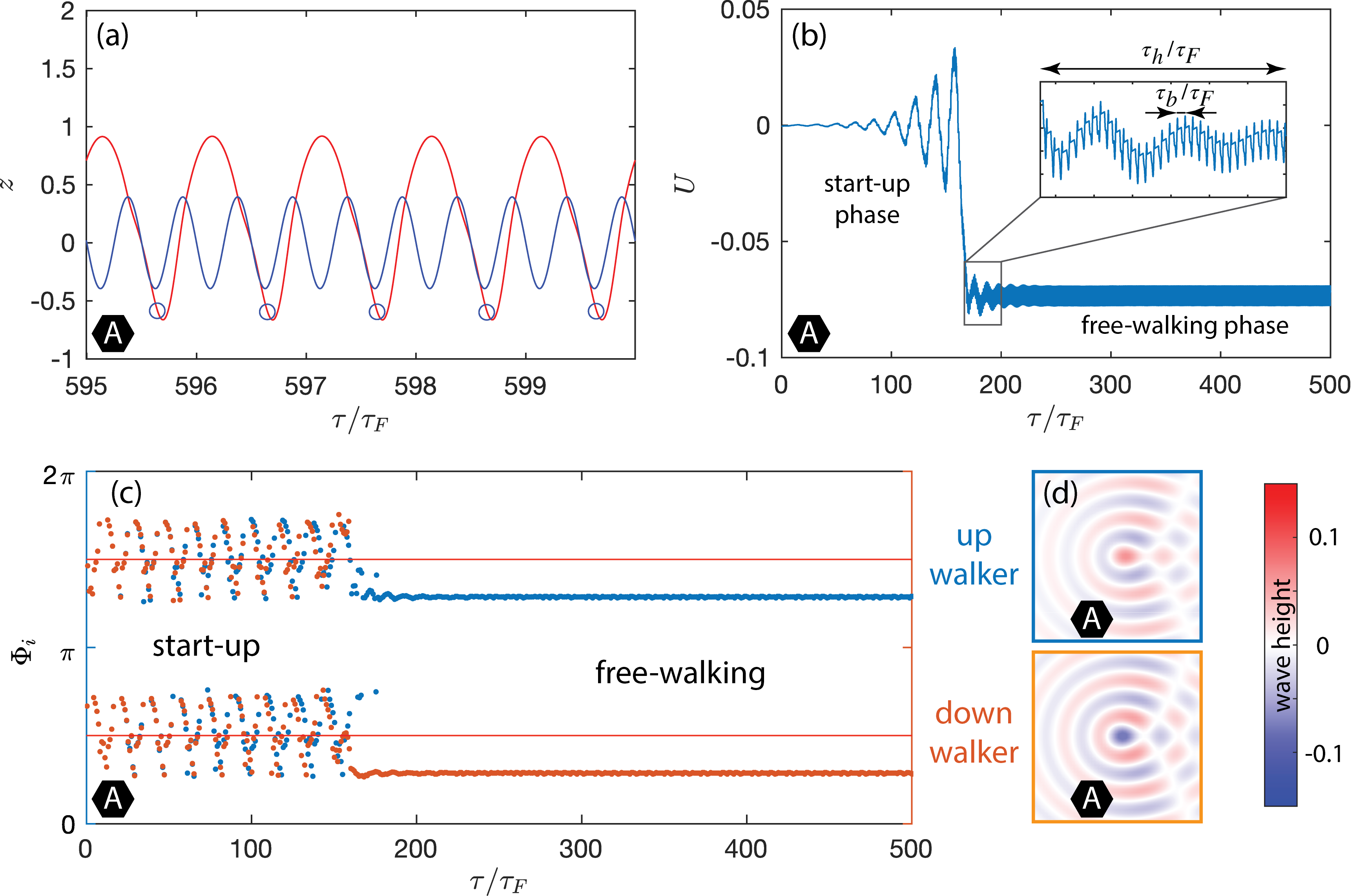}
    \caption{Simulations of non-resonant effects arising at the onset of motion of (2,1) walkers. (a)~Vertical trajectory of a (2,1) walker (red), for which the period of the droplet bouncing is twice the vibrational period of the fluid bath, commensurate with the Faraday period of its pilot wave. The blue curve represents the position of an unperturbed fluid bath, and the blue circles denote the location of the drop's base at impact time $\tau_i$. (b)-(c)~The start-up phase of the (2,1) walker for $\tau/\tau_F<170$ is characterized by two phenomena: periodic flipping of the impact phase and unstable lateral oscillations of the walker around its initial position. (b)~ The horizontal velocity profile of the walker reveals emergent features at two distinct time scales. After its startup phase, the drop exhibits underdamped speed oscillations over a timescale $\tau_h \sim 100 \tau_F$ before settling into a steady (2,1) walking state. The inset panel shows speed spikes associated with individual drop-bath impacts at the timescale $\tau_b \sim \tau_F$. (c)~Phase of drop impact $\Phi_i$ as defined in Eq.~\ref{eqn:h}. A walker can lock into one of the two stable walking states, with $\Phi_i$ either above or below $\pi$. Blue and orange dots correspond to walker simulations in the associated ``up'' and ``down'' states, respectively. (d)~The two walking states produce different strobed images of the wave field (color marks the wave height). Here, the wavefields were strobed at the impact time of the up walker. Simulations shown here correspond to point A ($\Omega=0.8$, $\Gamma/\Gamma_F=0.98$) in Fig.~\ref{fig:phase_diag}a.}
    \label{fig:up_down}
\end{figure}

We begin by exploring features of the coupled model with walkers exhibiting periodic vertical dynamics. As indicated in Fig.~\ref{fig:phase_diag}, the bouncing mode of a droplet is denoted by $(m,n)$ when the drop undergoes $n$ impacts in $m$ forcing periods of the bath. Notably, the stroboscopic pilot-wave models assume a resonant (2,1) mode for the drop. With the current model, stable (2,1) walkers emerge in the ($\Omega,\Gamma$) parameter space that is largely consistent with the experimental phase diagrams reported in Fig.~\ref{fig:phase_diag}. Fig.~\ref{fig:up_down}a shows a typical vertical trajectory of a (2,1) resonant walker, for which the period of the bouncing is twice that of the bath.

Each droplet impact excites a time-decaying standing wave with a form prescribed by Eq.~\eqref{eqn:h} that oscillates at half the frequency of the bath~\citep{benjamin_stability_1954,molacek_drops_2013b}. Therefore, one can calculate the impact phase with respect to a period of the pilot wave oscillations as
\begin{equation}
    \Phi_i = \frac{\int_{\tau_c} F_N(s)\frac{\Omega s}{2}ds}{\int_{\tau_c} F_N(s)ds} \hspace{2mm} (\text{mod }2\pi).
    \label{eqn:Phi}
\end{equation}
Notably $\Phi_i$ determines the state of the bath and the pilot wave field at the onset of drop impact. When $\Phi_i = \pi/2$ or $\Phi_i = 3\pi/2$, the drop impacts when the bath is at its peak upward velocity, when $h(\mathbf{x},\tau)$ is flat. At these phases, the largest momentum is transferred from the bath to the drop, but this has no horizontal component since $\nabla{h}$ vanishes. These impact phases that induce negligible lateral drop motion are indicated by red lines in Fig.~\ref{fig:up_down}c.

\textbf{Swaying onset of motion, up/down walkers}. A (2,1) bouncer starting from rest may sway back and forth along a line before reaching a stable state of rectilinear motion (see Fig.~\ref{fig:experiments}a and Video 1). This swaying onset of motion is most readily observed with bath oscillation frequency near $70$\;Hz (Fig.~\ref{fig:up_down}b). Here, the swaying motion of the walker is accompanied by the flipping of $\Phi_i$ between values that are above and below $\pi$, and periodic $\Phi_i$ variations between the flips (Fig.~\ref{fig:up_down}c). Eventually, the drop locks into one of the two possible walking states, specifically `up' or `down' states as indicated in blue or orange in Fig.~\ref{fig:up_down}c,d. These two walking states are readily observed in experiments (see green and red tracks in Video 4) and distinguished by the presence of a dark spot beneath the drop in one of the two states. Because stroboscopic models assume that the phase $\Phi_i$ is constant, they cannot distinguish between the up and down walking states. Likewise, the stroboscopic models cannot capture the swaying onset of motion, where flipping of the impact phase $\Phi_i$ is evidently important.

It is worth pointing out the two distinct timescales apparent in Fig.~\ref{fig:up_down}b. The first timescale ($\tau_b \sim \tau_F$), evident in the velocity spikes in Fig.~\ref{fig:up_down}b inset, characterizes the sharp velocity changes induced by impact. The second timescale ($\tau_h \sim \lambda_F/u_0$) characterizes the drop's speed modulations, specifically, the drop undergoes unstable speed oscillations before ($\tau<170\tau_F$), and underdamped speed oscillations immediately after ($170\tau_F<\tau<250\tau_F$), locking into its free walking state at speed $u_0$. 

\subsection{Exotic free walkers}

\begin{figure}
    \centering
    \includegraphics[width=.9\linewidth]{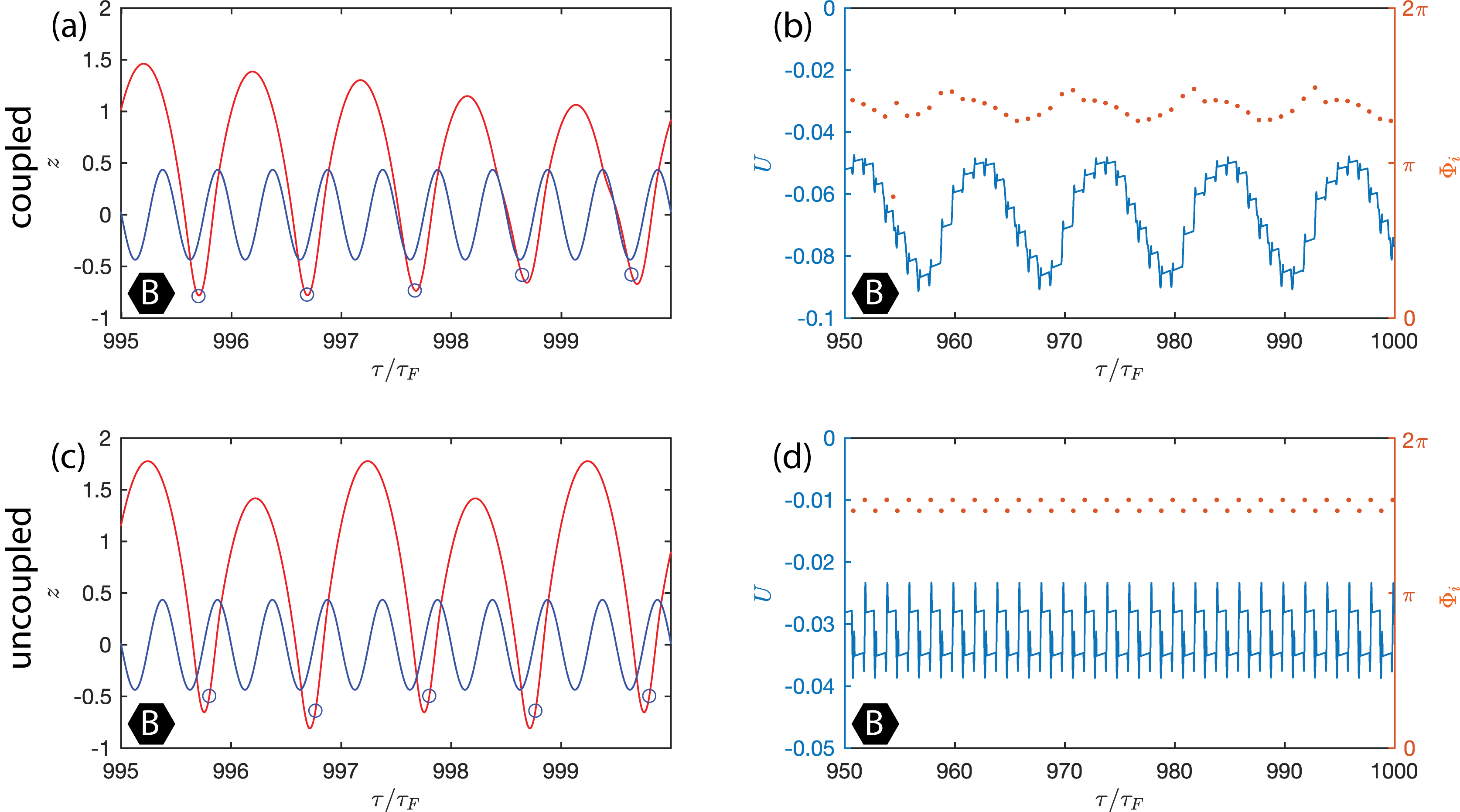}
    \caption{Simulations of the mixed-state walker corresponding to point B ($\Omega=0.7$, $\Gamma/\Gamma_F=0.99$) in Fig.~\ref{fig:phase_diag}a and the experiment shown in Fig.~\ref{fig:experiments}c. (a)~Vertical trajectory of the mixed state reveals that it is a (22,11) mode, which is very close to the (24,12) state reported in experiments~\citep{wind-willassen_exotic_2013}. (b)~Both speed and impact phase change periodically in time; the length scale of these oscillations is comparable to $\lambda_F$. If $h$ and $\dot{h}$ are neglected in Eq.~\eqref{eqn:FN}, the drop's vertical trajectory is no longer influenced by the wavefield and (c)~the mixed state relaxes to a (4,2) mode, (d)~in which $\lambda_F$-scale oscillations in speed and $\Phi_i$ are absent.}
    \label{fig:mixed_walker}
\end{figure}

\textbf{``Mixed-state'' walkers}. The two-way coupling implemented in our model also allows us to capture some of the more exotic walking states. For example, Fig.~\ref{fig:mixed_walker}a,b shows a ``mixed-state'' walker, for which both the drop speed and impact phase $\Phi_i$ fluctuate periodically. This state has been reported in the experiments of \citet{wind-willassen_exotic_2013} (see Video 3~\citep{wind-willassen_exotic_2013}), who hypothesized that the walker was switching periodically between the $(2,1)^1$ and $(2,1)^2$ walking modes (see Fig.~\ref{fig:phase_diag}). In our simulations, the speed oscillations emerge in a parameter regime consistent with that reported in the experiments of \citet{wind-willassen_exotic_2013} (see Video 5). Furthermore, the in-line speed oscillations have a wavelength corresponding to $\lambda_F$ and are sustained indefinitely for free walkers, features consistent with the speed changes evident in the walker trajectory shown in Fig.~\ref{fig:experiments}c. Experiments show that such walkers can transition into one of the (2,1) modes after collisions with boundaries. A similar transition may arise in our simulations when walkers are confined by a central force. 

We emphasize that the two-way wave-droplet coupling implemented in the current model is critical for capturing these ``mixed-state'' walkers. Specifically, the periodic speed oscillations of the free ``mixed-state'' walking vanish if one uses only the one-way coupling implemented in \citet{wind-willassen_exotic_2013}. Note that with only one-way coupling, $h$ and $\dot{h}$ terms are neglected in equation~\eqref{eqn:FN}, and the drop relaxes to a stable (4,2) walking state (Fig.~\ref{fig:mixed_walker}c,d).

\textbf{Intermittent walkers}. 
Both \citet{molacek_drops_2013a} and \citet{wind-willassen_exotic_2013} reported that a relatively large drop can exhibit intermittent walking marked by rectilinear motion and sporadic reversals in direction. Such intermittent walkers can be readily found experimentally (see Fig.~\ref{fig:experiments}b and Video~2). If one starts with a relatively large drop ($\Omega\approx0.9$, $R\approx0.41$\;mm) and gradually increases the driving acceleration, the drop transitions from a $(1,1)$ bouncer to a static chaotic bouncer to an intermittent walker. Further increasing the driving acceleration causes the drop to transition to a constant-speed $(2,1)^1$ mode. However, even in the latter state, the walker can temporarily transition back to an intermittent mode following collisions with boundaries. Note that this region of the phase diagram corresponds to the greatest mismatch between the experiments of \citet{wind-willassen_exotic_2013} and the $(m,n)$ state boundaries predicted by \citet{molacek_drops_2013a, molacek_drops_2013b}, which is likely due to the shortcomings of their assumption that $h(\mathbf{x},\tau)=0$ at impact in this high-memory parameter region.

\begin{figure}
    \centering
    \includegraphics[width=0.9\linewidth]{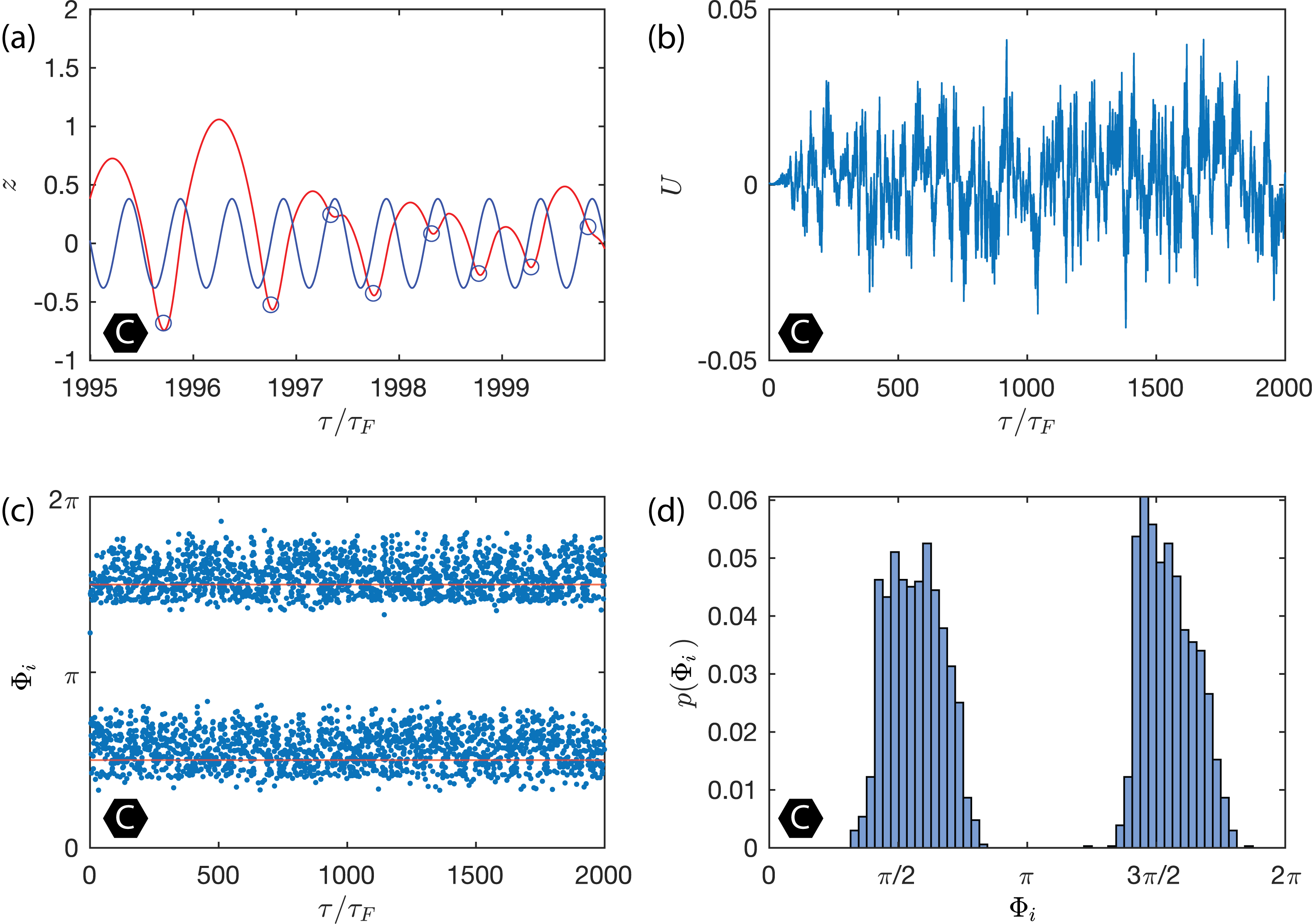}
    \caption{Simulation of the intermittent walker corresponding to point C ($\Omega=0.95$, $\Gamma/\Gamma_F=0.99$) in Fig.~\ref{fig:phase_diag}b and experiment in Fig.~\ref{fig:experiments}b. (a)~The drop's vertical dynamics is chaotic, and (b)~the instantaneous horizontal speed fluctuates sporadically about a zero mean. (c)~Time series and (d)~histogram of the impact phase $\Phi_i$ reveal its substantial variability about two dominant values.}
    \label{fig:intermittent}
\end{figure}

Intermittent walkers emerge in our model near the boundary between (2,1) walking and static chaotic states in the phase diagram (see Fig.~\ref{fig:phase_diag}b). Here, the vertical dynamics is chaotic (see Fig.~\ref{fig:intermittent}a), and the walker's horizontal speed fluctuates erratically around zero. The associated random fluctuations in $\Phi_i$ follow a bimodal distribution (Fig.~\ref{fig:intermittent}c,d). The centers of the two bands in the $\Phi_i$ distribution are separated by approximately $\pi$. Notably, owing to the substantial variability of $\Phi_i$, one expects this walking regime to be particularly poorly described by the stroboscopic models.

\begin{figure}
    \centering
    \includegraphics[width=\linewidth]{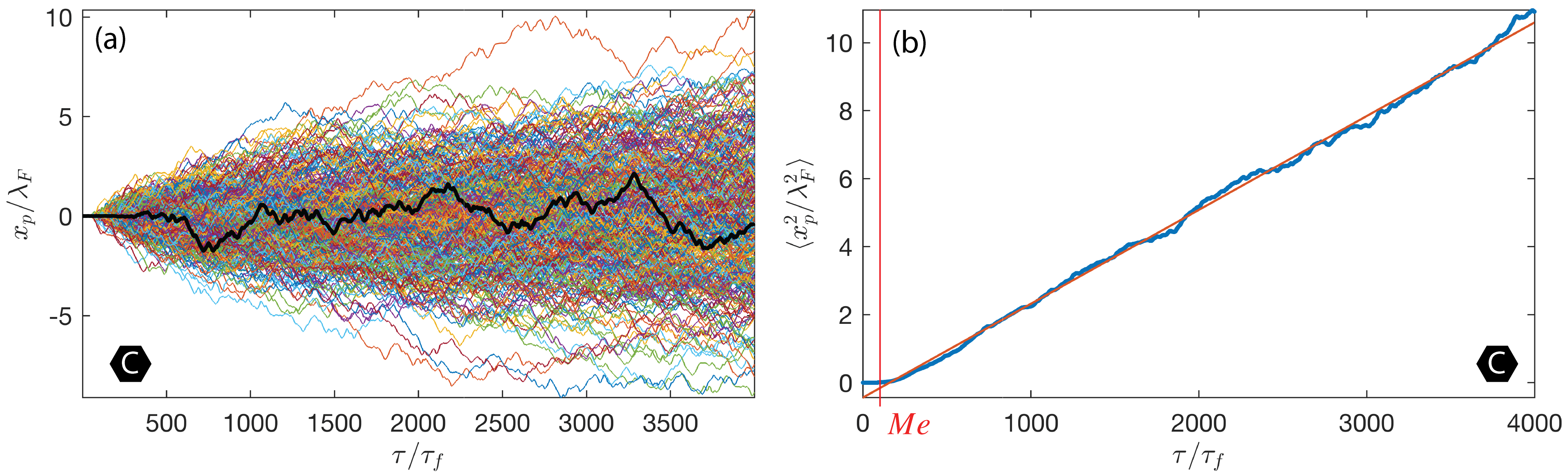}
    \caption{(a) Numerical simulation of 360 intermittent walkers moving along the $x$-axis, computed at point C ($\Omega=0.95$, $\Gamma/\Gamma_F=0.99$) in Fig.~\ref{fig:phase_diag}b. The black line shows a typical time evolution of an intermittent walker trajectory. (b) The evolution of the mean-square displacement of walkers in time is diffusive after an early-time transient of characteristic magnitude comparable to $\tau_F\mathit{Me}$. Here, $\langle x_p^2/\lambda_F^2 \rangle = 2 D \tau/\tau_F$ and $D=0.0014$ is the characteristic diffusion coefficient.}
    \label{fig:randomwalk}
\end{figure}

In Fig.~\ref{fig:randomwalk}a, we simulate three hundred and sixty intermittent walkers and show that their statistics exhibit features reminiscent of a random walk. All intermittent walkers were initially placed at $x=0$ with zero velocity. Then, a small perturbation to their speed was applied in the $x$-axis direction. This produced walker trajectories reminiscent of a classical one-dimensional random walker. The mean-square displacement of the walkers exhibited diffusive behaviour after a short transient period (of order $\tau_F\mathit{Me}$) in which intermittent walkers build up their wave field (Fig.~\ref{fig:randomwalk}b).

We note that the random walking state reported here is distinct from that achieved by \citet{tambasco_bouncing_2018} for walkers above the Faraday threshold. The random walkers described by \citet{tambasco_bouncing_2018} were influenced predominantly by interaction with the standing field of Faraday waves. Specifically, the drop navigated the background peaks and troughs at characteristic speed, $u_0$, changing direction after a characteristic distance $\lambda_F$, giving rise to a diffusion coefficient $D\approx\frac{\lambda_F u_0}{\omega_D R^2}\approx0.5$. Our experiments were performed below the Faraday threshold, where there is no standing Faraday wave to set the characteristic length scale of the random walk. Instead, the motion arises solely from the chaotic interactions of the walker with its pilot wave. This distinction is reflected in a significantly lower diffusion coefficient ($D\approx0.0014$). Moreover, intermittent walkers exhibit a predominantly one-dimensional random walk in the absence of ambient air currents, while walkers above the Faraday threshold diffuse in two dimensions.

\begin{figure}
    \centering
    \includegraphics[width=0.9\linewidth]{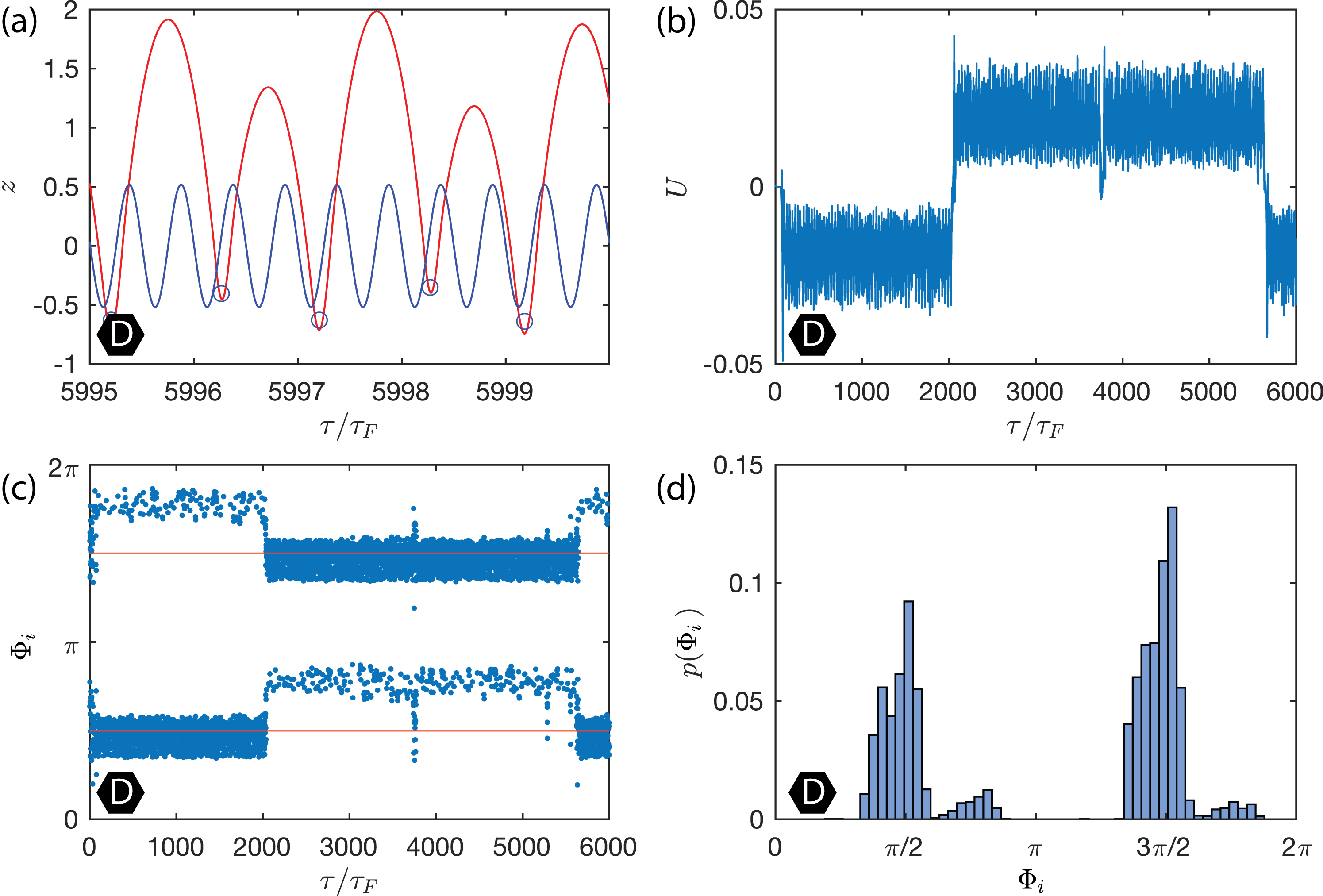}
    \caption{Simulations of a chaotic walker corresponding to point D ($\Omega=0.6$, $\Gamma/\Gamma_F=0.99$) in Fig.~\ref{fig:phase_diag}b. (a)~Vertical trajectory. Time evolution of (b)~horizontal speed and (c)~impact phase. Note that switching of the impact phase between $\Phi_i>\pi$ and $\Phi_i<\pi$ prompts the relatively rare changes in the walking direction over a reversal timescale $\tau_R \sim 1000\tau_F$. (d)~Histogram of the impact phase distribution.}
    \label{fig:chaotic}
\end{figure}

\textbf{Chaotic walkers}.
Chaotic walkers have been reported experimentally for relatively small drops at high memory~\citep{wind-willassen_exotic_2013} (see point D in Fig.~\ref{fig:phase_diag}b). While chaotic walkers (Fig.~\ref{fig:chaotic}) share many similarities with intermittent walkers (Fig.~\ref{fig:intermittent}), they have two distinguishing features. First, they maintain non-zero mean speed, so more closely resemble intermittent walkers with a drift. Second, they maintain noisy impact phase $\Phi_i$ near two values separated by $\pi$, with relatively rare switches between the two that correspond to the reversals in walking direction. The characteristic time scale of these reversals is $\tau_R\sim1000\tau_F$, at least three orders of magnitude larger than that of the intermittent walkers, for which $\tau_R\sim\tau_F$. We note that these phase flips are related to the ``time reversal effect'' reported in \citet{perrard_wave-based_2016}; however, they arise here spontaneously.

\section{Confined walkers} \label{sec:confined}
\begin{figure}
    \centering
    \includegraphics[width=\linewidth]{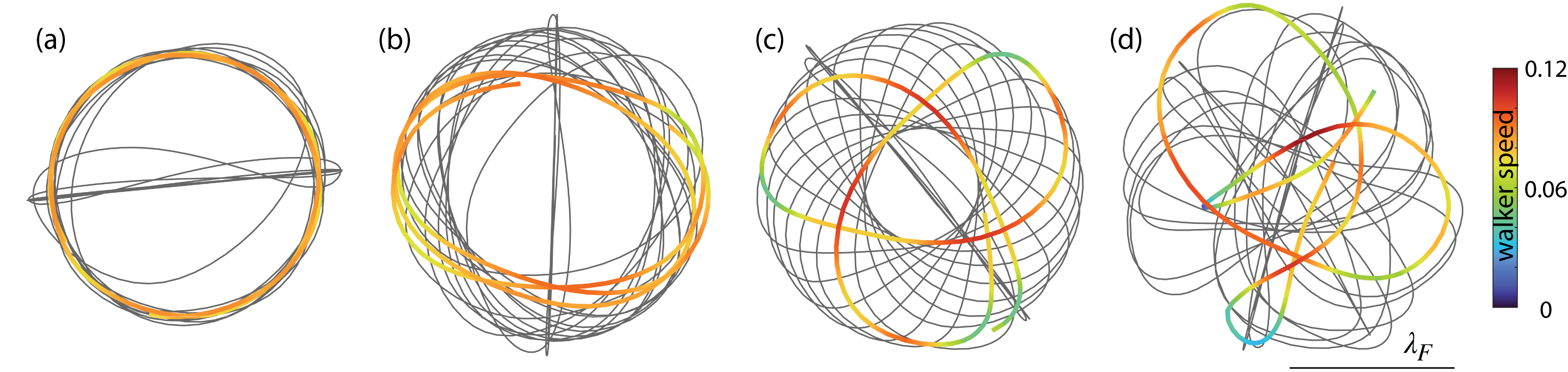}
    \caption{The trajectories of walkers confined by a dimensionless central force $F(r)=-7.85\cdot10^{-5}r$. (a)~Circular orbits at $\Gamma/\Gamma_F=0.955$. (b)~Drifting oval orbits at $\Gamma/\Gamma_F=0.97$. (c)~Drifting trefoil orbits at $\Gamma/\Gamma_F=0.98$. (d)~Chaotic horizontal trajectory at $\Gamma/\Gamma_F=0.99$. The simulations were performed at $\Omega=0.8$, $\Gamma_F=3.5$, $f=72\;$Hz. The trajectories are color-coded according to speed, revealing in-line speed oscillations with characteristic wavelength $\lambda_F$.}
    \label{fig:corral_traj}
\end{figure}

The variability of the walker impact phase apparent in some of the exotic walker states detailed above can also emerge at high $\mathit{Me}$ when resonant (2,1) walkers are placed in confinement, owing to the influence of the complex pilot wave field on the drop's vertical motion. In Fig.~\ref{fig:corral_traj} we confine a droplet, which would typically exhibit a stable (2,1) free walking state, and keep track of its $\Phi_i$ statistics. The confinement is generated by adding a radial central force, expressed in dimensionless form as $F(r)=-7.85\cdot10^{-5}r$, to the right-hand side of Eq.~\eqref{eqn:xdot}, where $r$ denotes the distance from the origin. As the system memory increases, the walker can display various periodic trajectories (Fig.~\ref{fig:corral_traj}a-c) before transitioning to chaotic horizontal dynamics at high $\mathit{Me}$ (Fig.~\ref{fig:corral_traj}d). Similar progressions have been reported experimentally for a walker in both the circular corral~\citep{harris_wavelike_2013, cristea-platon_walking_2018} and a simple harmonic potential~\citep{perrard_self-organization_2014}. 

\begin{figure}
    \centering
    \includegraphics[width=\linewidth]{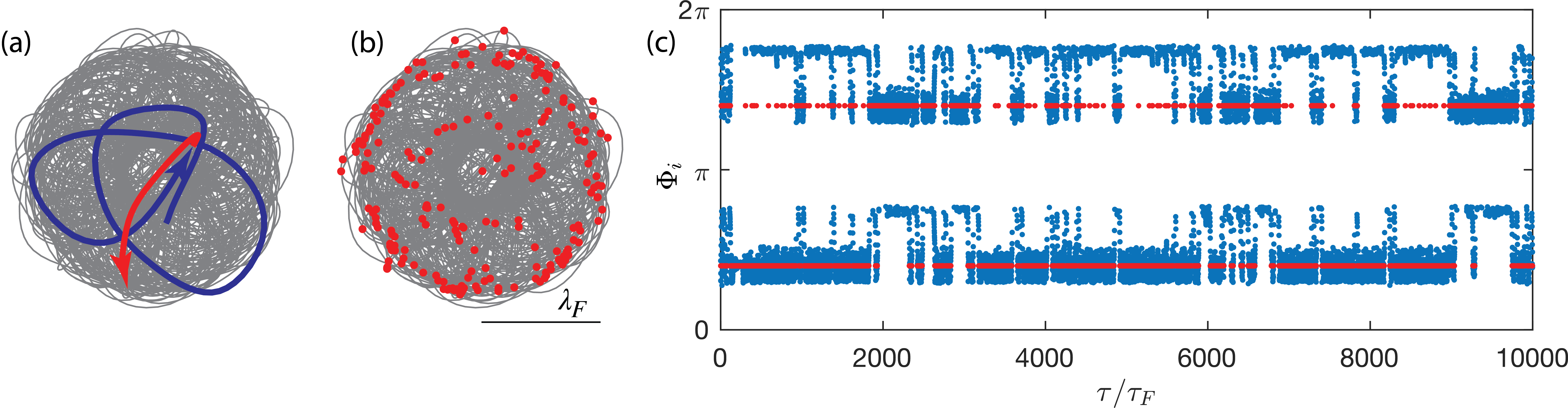}
    \caption{Simulation of the (2,1) walker detailed in Fig.~\ref{fig:up_down} when confined by a dimensionless central force $F(r)=-7.85\cdot10^{-5}r$ at high memory. (a)~The walker exhibits chaotic horizontal dynamics. The blue and red curves show a typical trajectory before and after the phase flip, respectively. (b)~Intermittent switches of the impact phase between up and down states, denoted by red dots, are distributed relatively uniformly within the domain, but arise most frequently at its outer edge. (c)~The walker's impact phase $\Phi_i$ is no longer constant and instead evolves similarly to that of the chaotic walker illustrated in Fig.~\ref{fig:chaotic}. The simulations were performed at $\Omega=0.8$, $\Gamma/\Gamma_F=0.99$, $\Gamma_F=3.5$, $f=72\;$Hz.}
    \label{fig:corral}
\end{figure}

In the simulation presented in Fig.~\ref{fig:corral}a a droplet follows a chaotic trajectory~\citep{cristea-platon_walking_2018}. While the emergence of the chaotic horizontal dynamics is captured in the high-memory limit of the stroboscopic models~\citep{durey_classical_2021}, our model captures a key feature that has been observed experimentally (Video 4~\citep{saenz_statistical_2017}), but not yet rationalized. Specifically, the walker exhibits intermittent switching between the up and down states described in \S\ref{sec:free_walkers}. The switching of the impact phase can be triggered when the drop changes direction in response to the confining central force or when it crosses its own path at high memory. In both cases, slowing of the walker transforms it into the early-time start-up phase of the walker shown in Fig.~\ref{fig:up_down}c (for $\tau/\tau_F<170$), where $\Phi_i$ flips intermittently between up and down states until it eventually locks into one or the other. Consequently, the possibility of sustained $\Phi_i$ switching arises. Fig.~\ref{fig:corral}a shows a typical walker trajectory before (blue) and after (red) the $\Phi_i$ switch, and Fig.~\ref{fig:corral}b plots the location of the phase switches during the course of 10000 Faraday periods. What was a stable (2,1) walker in the absence of confinement exhibits impact phase evolution reminiscent of a chaotic walker when confined (compare Fig.~\ref{fig:chaotic}c and Fig.~\ref{fig:corral}c).

Figs.~\ref{fig:chaotic} and \ref{fig:corral} both show evidence of walkers changing direction and retracing their path following flips between up and down states. The chaotic walker shown in Fig.~\ref{fig:chaotic} retraces its path upon the flip in $\Phi_i$, until another flip occurs. A strong correlation between $\Phi_i$ flips and the change in the direction of motion is evident upon comparison of Fig.~\ref{fig:chaotic}b and Fig.~\ref{fig:chaotic}c. The confined walkers of Fig.~\ref{fig:corral} respond similarly to phase flips. However, walkers retrace their previous path for a relatively short time owing to the relatively high memory in our simulations and the associated complexity of the guiding wavefield. A key difference between the simulations reported here and the induced phase flips examined in the experiments of \citet{perrard_wave-based_2016}, is that our flips emerge spontaneously due to the interactions of the walker's vertical dynamics with its wave field.

\section{Discussion and Conclusions}\label{sec:conclusions}
Following the development of Moláček and Bush's~\citep{molacek_drops_2013a, molacek_drops_2013b} model of pilot-wave hydrodynamics, the assumption of fixed $\Phi_i$ was quickly adopted in the subsequent stroboscopic models~\citep{oza_trajectory_2013, durey_faraday_2017, durey_classical_2021} and considerable effort was directed towards improving the fidelity of the wave model. Here, we have shown that retaining the simplified wave model of \citet{molacek_drops_2013b} while coupling the wave to the walker's vertical dynamics captures a rich dynamics of non-resonant effects inaccessible to stroboscopic models. Specifically, our model provides new rationale for the swaying onset of walker motion, intermittent and chaotic walking states~\citep{molacek_drops_2013b, wind-willassen_exotic_2013}, as well as the mixed-state walkers reported in the experiments of \citet{wind-willassen_exotic_2013}.

The model also offers a natural framework for studying the motion of the walkers over a standing wave, as may arise when the bath is driven above threshold $\gamma>\gamma_F$ in some limited spatial domain, and which may play the role of an applied potential. Such a situation was considered by \citet{tambasco_exploring_2018}, who investigated a walker interacting with the standing Faraday wave induced above a relatively deep well. One can incorporate the influence of a standing wave by adding a forcing term to Eq.~\eqref{eqn:xdot}
\begin{equation*}
    \mathbf{\ddot{x}}_p(\tau) +(\mathcal{D}_h F_N(\tau) + \frac{9}{2}\mathit{Oh}_a) \mathbf{\dot{x}}_p(\tau) = - F_N(\tau) \nabla{(h(\mathbf{x}_p,\tau)} + \phi(\mathbf{x}_p)\cos{{\Omega \tau}/{2}}),
\end{equation*}
where $\phi(\mathbf{x})$ is the envelope of a standing wave oscillating at the Faraday frequency. This formulation makes clear that one may decompose the wave field into self-induced and standing wave components: a decomposition that will be exploited in upcoming HQAs, including a hydrodynamic analog of the Kapitza-Dirac effect~\citep{primkulov_diffraction_2024}. Such a formulation highlights the significance of the two impact phases identified here: walkers in the up state will effectively encounter a standing wave with the opposite sign of that encountered by the down state.

The model presented here highlights the prevalence of non-resonant walking states at high memory, where complex walker-wave interactions may result in stochastic-like evolution in impact phase $\Phi_i$. These deviations of $\Phi_i$ from the fixed value of resonant walkers have long been deemed responsible for the mismatch between experiments and theoretical predictions of the stroboscopic models in several pilot-wave hydrodynamics settings~\citep{bush_hydrodynamic_2021}, including orbital stability~\citep{harris_droplets_2014, oza_pilot-wave_2014, oza_pilot-wave_2014-1} and the stability of orbiting~\citep{oza_orbiting_2017}, ratcheting~\citep{galeano-rios_ratcheting_2018} and promenading~\citep{arbelaiz_promenading_2018} pairs. Moreover, there is ample evidence that non-resonant effects play a critical role in the emergent statistics; for example, phase-switching is readily apparent in the corral experiments of \citet{harris_wavelike_2013} and \citet{saenz_statistical_2017}. This study has made clear how such non-resonant effects arise and may influence both the walker dynamics and statistics.

While the current model has fair computational efficiency, by construction it requires that one resolve the drop-wave dynamics on the time scale $\tau_F$ of individual walker bounces. At high memory, our model's computational time increases linearly with $\mathit{Me}$: simulating 27 seconds of a walker in confinement in real time takes about 24 minutes of simulation time with an Apple M2 processor when $\mathit{Me}=800$ (or $\Gamma/\Gamma_F=0.999$). Nevertheless, one can in principle bridge the three relevant timescales of walker dynamics with our model, specifically those of bouncing ($\tau_b$), horizontal motion ($\tau_h$), and statistical convergence ($\tau_\text{stat}$). 

Our model also introduces the possibility of more efficient, `stochastic stroboscopic' models that capture the erratic evolution of the impact phase $\Phi_i$. For example, one could model the evolution of the mean $\Phi_i$ of the walker inside the corral shown in Fig.~\ref{fig:corral} as a Markov process governing the switching between up and down walker states. Then, the simulation data from Fig.~\ref{fig:corral}c would allow one to define a Markov transition matrix between drop bounces as
\begin{equation*}
    \mathbf{M}=\begin{pmatrix}
    0.97 & 0.03 \\
    0.02 & 0.98
\end{pmatrix}, 
\end{equation*}
where the walker has a 97\% chance of remaining in an up state, a 3\% chance of switching from up to down, a 2\% chance of switching from down to up, and a 98\% of remaining in a down state. By assuming that the phase switching has this Markovian structure, the complex evolution of $\Phi_i$ could thus be pre-computed and incorporated into a stroboscopic model yielding an efficient model that effectively captures the influence of phase switching.

Overall, our model offers a computationally efficient framework for exploring the wide range of pilot-wave-hydrodynamic phenomena rooted in non-resonant effects. This sets the stage for the development of a new generation of pilot-wave hydrodynamic models capable of resolving the system evolution over a statistical timescale while incorporating the non-resonant effects critical to both the dynamics and emergent statistics.

\textbf{Acknowledgements}. The authors gratefully acknowledge financial support from NSF CMMI-2154151, and valuable discussions with Jan Molacek, Carlos Antonio Galeano Ríos, Pedro Sáenz, Kyle McKee, Tino Damiani, and Andres Arroyo. Dedicated to our friend Peter Baddoo; your gentle presence and playful humour are deeply missed.

\bibliography{references.bib}

\end{document}